\documentclass[aps,prl,showpacs,twocolumn,superscriptaddress,amsmath,amssymb]{revtex4}

\usepackage{epsfig,afterpage}
\usepackage{graphicx}
\usepackage{amsfonts}
\usepackage{amssymb}
\usepackage{indentfirst}
\usepackage{amsmath,amsthm}
\usepackage{dsfont}

\usepackage{epsfig}
\usepackage{subfigure}

\usepackage{multirow}
\usepackage{pstricks}
\usepackage{lscape}
\usepackage{psfrag}

\usepackage{wasysym} 

\def\beq{\begin{equation}}
\def\eeq{\end{equation}}
\def\bea{\begin{eqnarray}}
\def\eea{\end{eqnarray}}
\def\bq{\begin{quote}}
\def\eq{\end{quote}}
\def\ben{\begin{enumerate}}
\def\een{\end{enumerate}}
\def\bit{\begin{itemize}}
\def\eit{\end{itemize}}
\def\nn{\nonumber}

\begin{document}

\title{Matrix Product States for dynamical simulation of infinite chains}

 \author{M. C. Ba\~nuls}\email{banulsm@mpq.mpg.de} 
 \affiliation{Max-Planck-Institut f\"ur Quantenoptik,
 Hans-Kopfermann-Str. 1, 85748 Garching, Germany.}
 \author{M. B. Hastings}
 \affiliation{Microsoft Research, Station Q, CNSI Building, University of California, 
Santa Barbara, CA, 93106.}
\author{F. Verstraete}
\affiliation{Fakult\"at f\"ur Physik, Universit\"at Wien, Boltzmanngasse 5, A-1090 Wien, Austria.}
 \author{J. I. Cirac}
 \affiliation{Max-Planck-Institut f\"ur Quantenoptik,
 Hans-Kopfermann-Str. 1, 85748 Garching, Germany.}

\date{\today}

\begin{abstract}
We propose a new method for computing the ground state properties and
the time evolution of infinite chains based on a transverse contraction
of the tensor network. The method does not require finite size
extrapolation and avoids explicit truncation of the bond dimension
along the evolution. By folding the network in the time direction
prior to contraction, time dependent expectation values and dynamic
correlation functions can be computed after much longer evolution time
than with any previous method. Moreover, the algorithm we propose can
be used for the study of some non-invariant infinite chains, including impurity
models.
\end{abstract}


\maketitle

\begin{figure}[floatfix]
\psfrag{evol}[tc][tc]{(1)}
\psfrag{trun}[tc][tc]{(2)}
\psfrag{exp}[tc][tc]{(3)}
\includegraphics[width=\columnwidth]{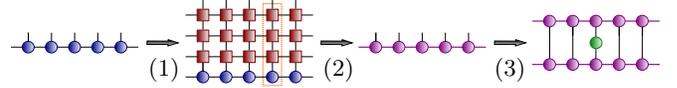}  
\caption{Standard time evolution with MPS. We start with a MPS state, represented here by a chain
of circles (tensors) connected by summed indices, with open lines for the physical spin indices.
On this state, a sequence of MPO is applied for each step of evolution (1), and the result is truncated (2)
to the maximal bond dimension $D$. After iterating the evolution, expectation values are computed (3) in the 
final state.
}
\label{fig:tn-a}
\end{figure}
Numerical simulation has become a fundamental tool to study quantum
many-body systems in condensed matter physics.  Unfortunately, the
exponential scaling of the dimension of the Hilbert space with system
size means that brute-force methods are only practical for very small
system sizes.  
However, other techniques based on Matrix Product
States (MPS)~\cite{aklt88,kluemper91,fannes92fcs,verstraete04dmrg,perez07mps}, 
\nocite{kluemper92} 
such as DMRG~\cite{white92dmrg}, achieve excellent results in one
dimension.
\nocite{schollwoeck05dmrg}



The success of DMRG methods is based on the fact that many interesting
physical states, including the ground states of many local Hamiltonians, can be
well approximated by a MPS. For a chain of $N$ $d$-dimensional
systems, this has the form 
\beq 
|\Psi\rangle =\sum_{i_1,\ldots
  i_N=1}^d \mathrm{tr}(A_1^{i_1}\ldots A_N^{i_N}) |i_1,\ldots i_N
\rangle.
\label{eq:mps}
\eeq
Each $A_k^i$ is a $D$-dimensional matrix.  
An important feature of MPS is that
expectation values of local operators can be efficiently computed,
allowing them to be used variationally.  
DMRG excels in the computation of static properties
of finite chains with local
Hamiltonians, where the required bond dimension grows slowly with the
size of the system~\cite{verstraete06mpsgs,hastings07area}, and of translationally invariant infinite
chains~\cite{ostlund95td,vidal07infinite}.  However, these methods can break down in time-dependent
problems far from equilibrium and also encounter difficulties in
dealing with infinite chains containing impurities.

MPS algorithms for non-equilibrium dynamics~\cite{cazalilla02tdmrg,white04realt,daley04adapt,verstraete04mpdo,vidal03eff} work well when the
system is close to its ground state, but when the system is far from
equilibrium the entanglement entropy may grow linearly in time and the
dimension $D$ required to describe the system will grow 
exponentially~\cite{calabrese05,schuch08entropy,osborne06efficient}, causing these methods to break down.  Although
improved algorithms have been developed based on finite propagation
speed of correlations~\cite{hastings08lightcone}, all known methods are limited to special cases or short
times.



Here we propose an alternative method to compute both ground states
and dynamical quantities for infinite chains within the MPS
formalism. Based on the transverse contraction of the
tensor network, it allows the study of problems not accessible by 
other methods, such as an impurity in an infinite system.
It enables the calculation of time-dependent expectation
values of local observables and of few body correlation
functions at different times, in a new much simpler way.
Moreover, by a folding of the network described below, the new method enables dynamical
studies that range much further in time than any other existing
method. 


The standard way of computing dynamical 
quantities with the MPS
formalism starts with a state that is (exactly)
described by a MPS (\ref{eq:mps}).  
Then, some evolution operator is applied to it for a given time,
making use of a Suzuki-Trotter expansion~\cite{trotter59}
\nocite{suzuki90}
of the total evolution operator.  
Within each discrete time step,
the evolution operator is broken down into a
product of operators.
In particular, for a nearest neighbor Hamiltonian $H=\sum_i h_{i,i+1}$,
we may write 
$ e^{-iH\delta}\approx e^{-i H_e \delta/2}e^{-i H_o\delta}e^{-i H_e
  \delta/2}, $ where $H_e$ ($H_o$) contains the $h_{i,i+1}$ terms with
even (odd) $i$, so that each exponential factor is a product of
mutually commuting local terms. 
Alternatively, the evolution operator can be decomposed as a product
of translationally invariant Matrix Product Operators
(MPO)~\cite{murg08mpo}.  The action of one step of evolution on the
MPS can be computed by applying the corresponding sequence of
operators (Fig.~\ref{fig:tn-a}) to yield a MPS with larger bond
dimension.  This must be truncated to keep the best MPS description of
the evolved state with fixed 
dimension, $D$. After repeating this procedure for the required number
of steps, expectation values can be calculated in the evolved state.
The accuracy of the description will however drop exponentially with
the successive truncations. 

\begin{figure}[floatfix]
\hspace{-.05\columnwidth}
\begin{minipage}[c]{.4\columnwidth}
\subfigure[Transverse contraction along space direction renders a finite 2D network.]{
 \label{fig:tn-b}
\psfrag{contractL}[c][c]{contraction}
\psfrag{contractR}[c][c]{}
\psfrag{Eo}[c][]{$E_O$}
\psfrag{E}[c][]{$E$}
\psfrag{langle}[c][]{$\langle L |$}
\psfrag{rangle}[c][]{$|R\rangle$}
\psfrag{time}[bc][bc]{time}
\psfrag{space}[tc][tc]{space}
  \includegraphics[height=.9\columnwidth]{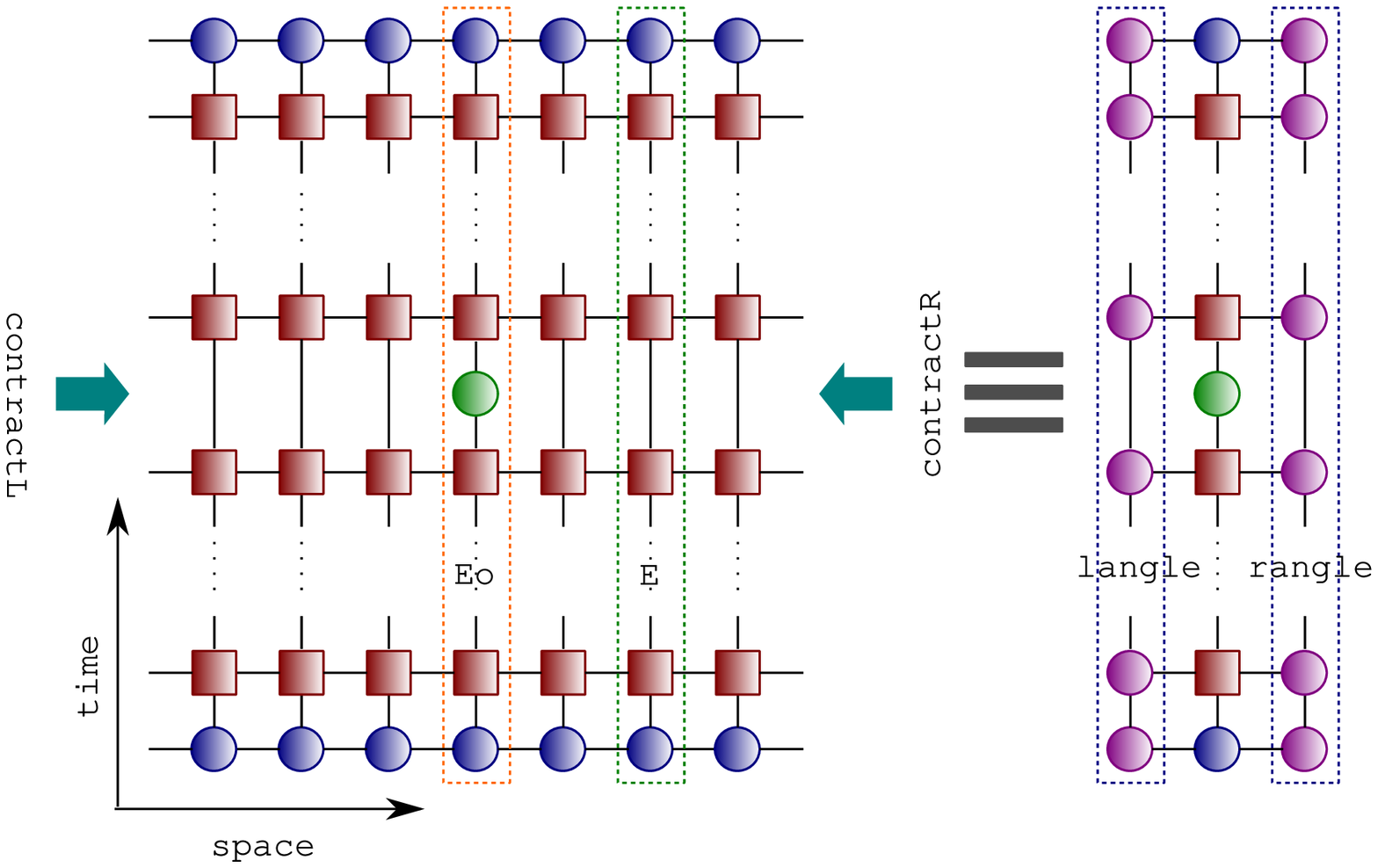}
} 
\end{minipage}
\hspace{.185\columnwidth}
\begin{minipage}[c]{.4\columnwidth}
\subfigure[Transverse  contraction of folded network.]{
 \label{fig:tn-fold}
\psfrag{Eo}[bc][bc]{$\tilde{E}_O$}
\psfrag{E}[bc][bc]{$\tilde{E}$}
\psfrag{langle}[c][c]{$\langle \tilde{L} |$}
\psfrag{rangle}[c][c]{$|\tilde{R}\rangle$}
\psfrag{fol}[cl][cl]{\parbox{.2\columnwidth}{folding axis}}
  \includegraphics[height=.82\columnwidth]{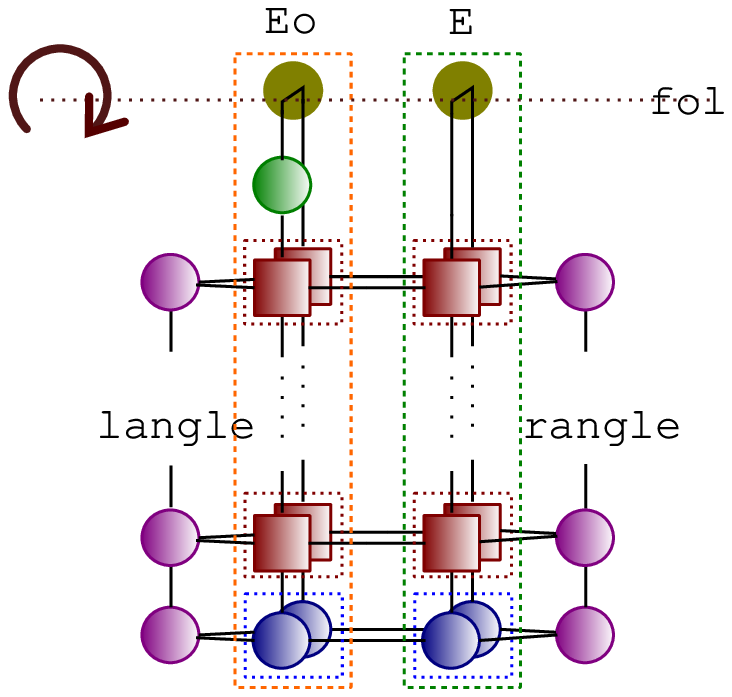}
}
\end{minipage}
\caption{Expectation value $\langle O(t)\rangle$ with the basic transverse 
method (a) and with folding (b), where operators for the same time 
step are grouped together in a double effective operator.
}
\end{figure}

Our new method avoids this explicit truncation on the
bond dimension of the evolved MPS.  The basic idea is to look at the
quantity that we want to compute, say the time dependent expectation
value of some local operator, $\langle \Psi(t)|O|\Psi(t)\rangle$, as
the contraction of a two dimensional tensor network, and perform it, 
not along time, but in the direction of space (see Fig.~\ref{fig:tn-b}).  To
construct the network, we start from the initial MPS and, for every
evolution step, apply the proper MPOs.  Repeating this for the
required number of evolution steps, we construct  
the exact evolved MPS (within the Trotter approximation), as 
no truncation is carried out. 
Finally, we apply the local operator $O$
and contract with the Hermitian conjugate of
the evolved state as constructed before.

The procedure above produces a two dimensional network, infinite in the
spatial direction as the original MPS, but finite along the time
direction. 
The expectation value we want to
compute can now be written as~\cite{perez07mps}, 
$$
\langle \Psi(t)|O|\Psi(t)\rangle =
\lim_{k\rightarrow\infty}
\mathrm{tr}(E_{\ell}^{[-k]} 
\ldots E^{[-1]} E_O^{[0]} E^{[1]}\ldots
E_{r}^{[k]}), $$
where 
$E(t)=\sum_i {\bar{A}^i}(t)\otimes A^i(t)$
is the transfer matrix of the evolved state,  
$E_O(t)=\sum_{i,j}[{\bar{A}^i}(t)\otimes A^j(t)] \langle i|O|j\rangle$
contains the only application of the single-body operator,  
and the
bracketed superindices on each transfer matrix indicate the site of
the chain.  For a translationally invariant MPO
representation of the evolution
operator, the network retains the
invariance~\footnote{Using the
  standard decomposition of the evolution operator into commuting
  products, the resulting network has translational
  symmetry with period two, so that the argument can be easily adapted
  substituting $E$ for the product of two contiguous transfer
  matrices, $E_e E_o$.} and the transfer matrix is the same on every
site, except for the single one on which $O$ acts~\footnote{In the infinite limit
  we do not need to consider the vector terms at the edges, $E_{\ell}^{[-k]}$ and
  $E_{r}^{[k]}$, which would also be different for the finite case.}. 
If the largest eigenvalue of $E(t)$, $\lambda$, is non-degenerate,
$ E^k(t)\xrightarrow[k \rightarrow \infty]{}\lambda^k|R\rangle\langle
L| $. 
Effectively, we may then substitute the left and right semi-infinite lattices at both
sides of the operator by the left and right eigenvectors of $E(t)$ 
corresponding to the largest eigenvalue, $\langle L|$
and $|R \rangle$, 
\beq
\langle O(t) \rangle=\frac{\langle \Psi(t)|O|\Psi(t)\rangle}{\langle \Psi(t)|\Psi(t)\rangle}
=\frac{\langle L | E_O | R \rangle}{\langle L | E | R \rangle}.
\label{eq:O(t)}
\eeq

We now specify the algorithm for computing time-dependent expectation
values in translationally invariant infinite chains.
The first step is to find the best MPS
approximation, with given bond dimension $D$, to the dominant
eigenvectors of $E(t)$.  
To this end, we repeatedly apply
the transfer matrix $E(t)$ (already written as a MPO
along the time direction, see Fig.~\ref{fig:tn-b})
to the left and to the right of an arbitrary
initial MPS vector and truncate the result to the chosen $D$,
using the technique for two dimensional tensor 
networks introduced in~\cite{verstraete04mpdo,murg07hard}, 
until convergence is achieved.
The procedure yields a MPS approximation to the eigenvectors, with the
truncation taking always place in the space of transverse vectors.
The second step, computing the numerator and the denominator in (\ref{eq:O(t)}), 
can be done very efficiently, as each term is a contraction of a MPO 
acting between a pair of MPS.
The adaptation of the method to the case of imaginary time evolution
is straightforward, so that it is also useful for finding ground state
properties. 
In this case, the eigenvector calculation is similar to that in
transfer matrix DMRG algorithms for thermal states~\cite{bursill96trans}.
\nocite{wang97trans}

\begin{figure}
\hspace{-.05\columnwidth}
\begin{minipage}[c]{.3\columnwidth}
\subfigure[Tensor network corresponding to the magnetization at distance $x$ from the impurity.]{
 \label{fig:impSch}
\psfrag{i0}[c]{$i=0$}
\psfrag{ix}[c]{$i=x$}
\psfrag{langle}{$\langle L |$}
\psfrag{rangle}{$|R\rangle$}
\includegraphics[height=1.2\columnwidth]{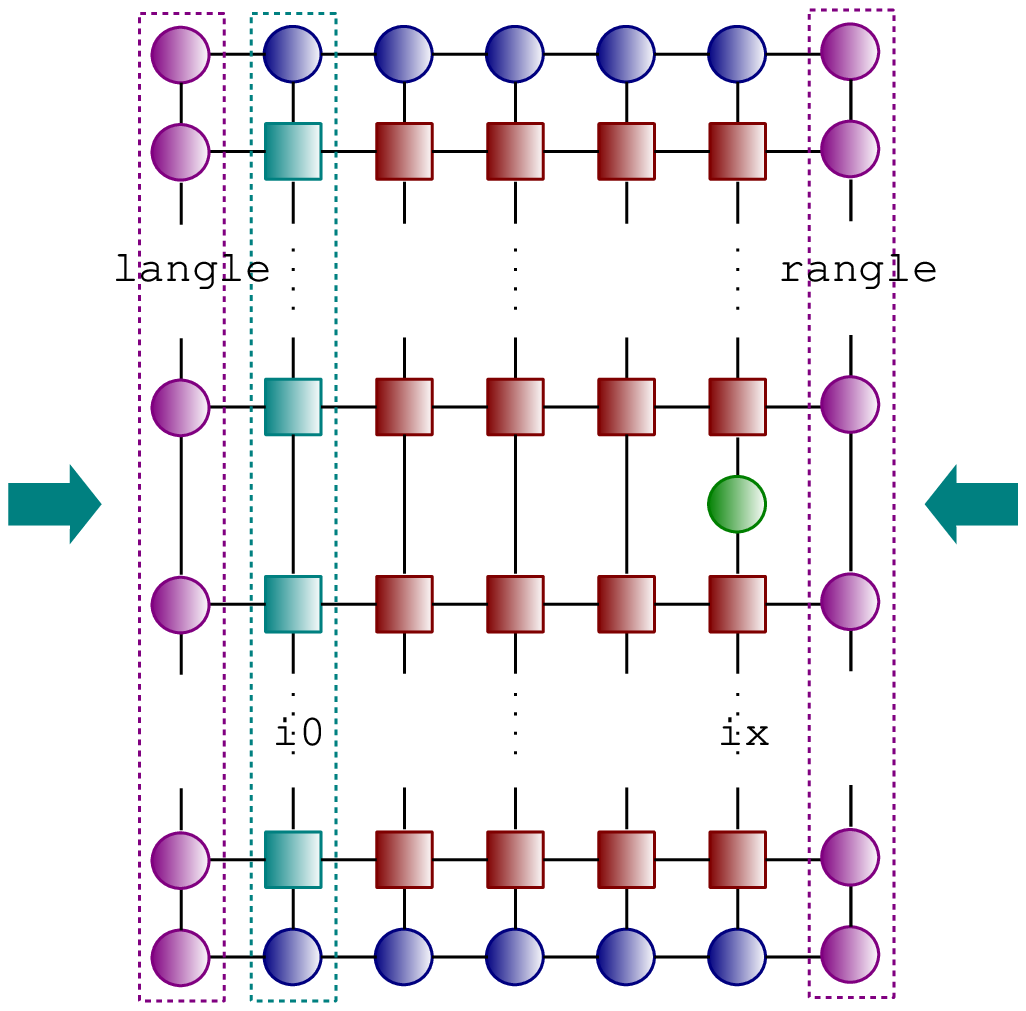}
} 
\end{minipage}
\hspace{.05\columnwidth}
\begin{minipage}[c]{.6\columnwidth}
\subfigure[$\langle \sigma_x\rangle$ as a function of distance $x$.]{ 
 \label{fig:impPlot}
\begin{psfrags}%
\psfragscanon%
%
\psfrag{s04}[][]{\color[rgb]{0,0,0}\setlength{\tabcolsep}{0pt}\begin{tabular}{c} \end{tabular}}%
\psfrag{s05}[][]{\color[rgb]{0,0,0}\setlength{\tabcolsep}{0pt}\begin{tabular}{c} \end{tabular}}%
\psfrag{s12}[t][t]{\color[rgb]{0,0,0}\setlength{\tabcolsep}{0pt}\begin{tabular}{c}$g_0=1$\end{tabular}}%
\psfrag{s13}[t][t]{\color[rgb]{0,0,0}\setlength{\tabcolsep}{0pt}\begin{tabular}{c}$g_0=2$\end{tabular}}%
\psfrag{s14}[t][t]{\color[rgb]{0,0,0}\setlength{\tabcolsep}{0pt}\begin{tabular}{c}$g_0=0.5$\end{tabular}}%
%
\psfrag{x01}[t][t]{0}%
\psfrag{x02}[t][t]{0.1}%
\psfrag{x03}[t][t]{0.2}%
\psfrag{x04}[t][t]{0.3}%
\psfrag{x05}[t][t]{0.4}%
\psfrag{x06}[t][t]{0.5}%
\psfrag{x07}[t][t]{0.6}%
\psfrag{x08}[t][t]{0.7}%
\psfrag{x09}[t][t]{0.8}%
\psfrag{x10}[t][t]{0.9}%
\psfrag{x11}[t][t]{1}%
\psfrag{x12}[t][t]{-30}%
\psfrag{x13}[t][t]{-20}%
\psfrag{x14}[t][t]{-10}%
\psfrag{x15}[t][t]{0}%
\psfrag{x16}[t][t]{10}%
\psfrag{x17}[t][t]{20}%
\psfrag{x18}[t][t]{30}%
%
\psfrag{v01}[r][r]{0}%
\psfrag{v02}[r][r]{0.1}%
\psfrag{v03}[r][r]{0.2}%
\psfrag{v04}[r][r]{0.3}%
\psfrag{v05}[r][r]{0.4}%
\psfrag{v06}[r][r]{0.5}%
\psfrag{v07}[r][r]{0.6}%
\psfrag{v08}[r][r]{0.7}%
\psfrag{v09}[r][r]{0.8}%
\psfrag{v10}[r][r]{0.9}%
\psfrag{v11}[r][r]{1}%
\psfrag{v12}[r][r]{0.4}%
\psfrag{v13}[r][r]{0.5}%
\psfrag{v14}[r][r]{0.6}%
\psfrag{v15}[r][r]{0.7}%
\psfrag{v16}[r][r]{0.8}%
\psfrag{v17}[r][r]{0.9}%
\psfrag{v18}[r][r]{1}%
%
\includegraphics[height=.8\columnwidth,width=.9\columnwidth]{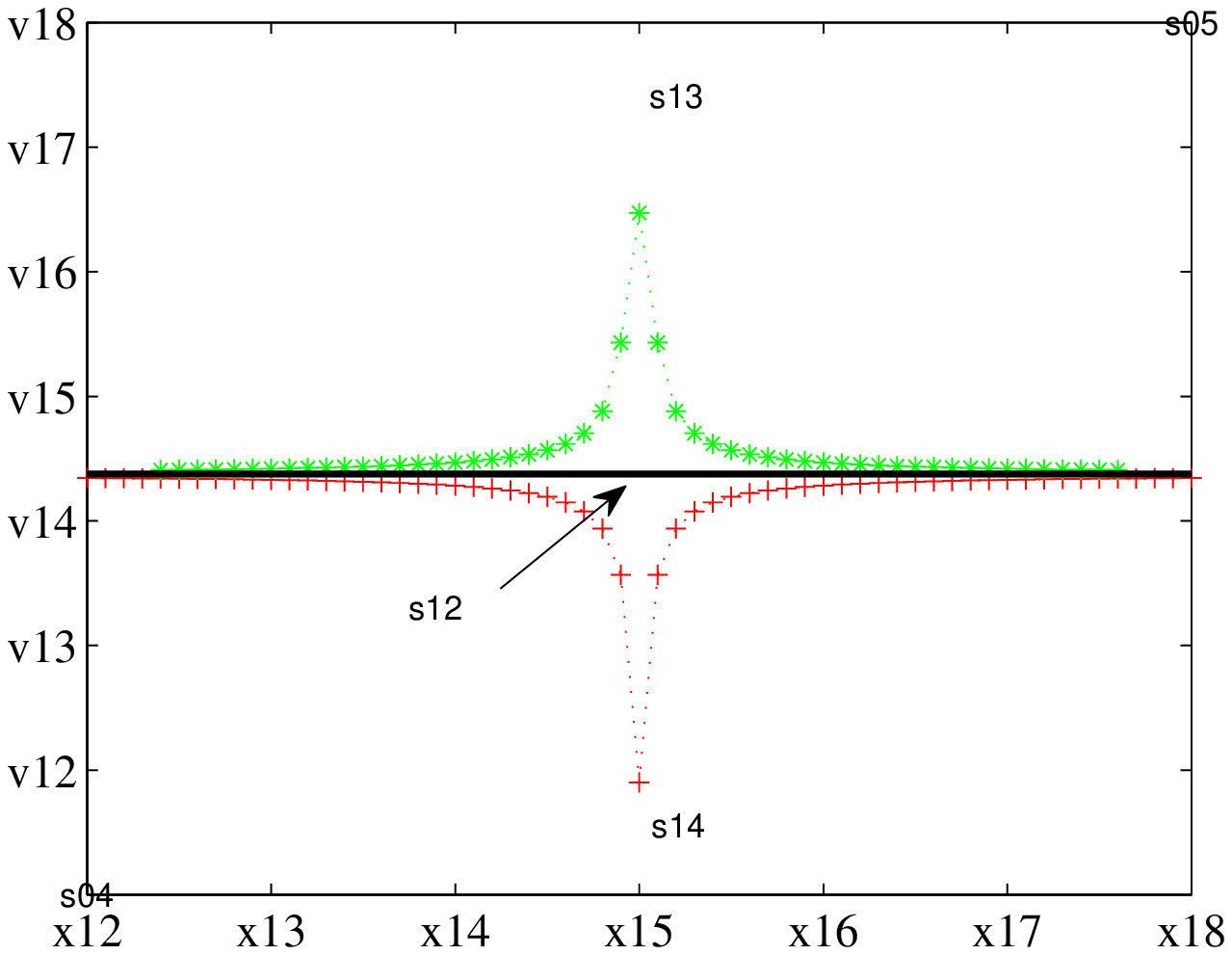}
\end{psfrags}
}
\end{minipage}
\caption{Ising chain with magnetic impurity at the origin.
}
 \label{fig:impurity}
 \end{figure}

With this approach, we study an infinite chain with an impurity. 
We consider an Ising chain,
\beq
H=-\left(\sum_i\sigma_z^i \sigma_z^{i+1}+g_i \sigma_x^i \right),
\label{eq:Ising}
\eeq
with $g_i=1$ $\forall i\neq 0$, and the impurity represented by a different value of the field at site $i=0$, $g_0$.
The system is started in a product MPS
and imaginary time evolution is applied for a long time, so that we
approach the ground state.  
Then we compute the site dependent magnetization, $\langle \sigma_x^{[i]} \rangle$.
Such a calculation cannot be
easily done with a purely invariant method as iTEBD~\cite{vidal07infinite}, because the
presence of a singular site will affect a cone of tensors as time
increases.  However, with the transverse method, the 
computation of $\langle L |$ and $|R\rangle$ is not modified by
the presence of the impurity~\cite{rommer99imp}.
 Thus the cost of computing the expectation value of a
local operator acting on the position $i=0$ in the ground state of this chain will
be the same as in the translationally invariant case, while applying 
the operator at $i\neq 0$ will reduce to the contraction of a 2D tensor network
of width $i+3$  (Fig.~\ref{fig:impurity}).

The capabilities of the transverse method regarding real time
evolution can be further illustrated by the computation of two-body
correlators at different times. If we consider two different times, $t_2>t_1$, we may
write 
\bea 
\langle &\Psi(0)&|O_2^{[x]}(t_2) O_1^{[x+\Delta]}(t_1)
|\Psi(0)\rangle \nn \\ &=&\langle \Psi(0) | U(t_2,0)^{\dagger}
O_2^{[x]} U(t_2,t_1) O_1^{[x+\Delta]} U(t_1,0) |\Psi(0)\rangle \nn
\\ &=&\frac{\langle L | E_{O_2(t_2)} E^{\Delta-1} E_{O_1(t_1)} | R
  \rangle}{\langle L | E^{\Delta+1} | R \rangle},
\label{eq:O(t)O(t)}
\eea 
where $E$ is the transfer matrix resulting from evolution until
time $t_2$, and $E_{O_i(t)}$ are the corresponding MPO containing the
action of each single-body operator~\cite{naef99}. 

In particular, if both operators act on the same site ($\Delta=0$), 
the computation has the same cost as one single expectation value.
If $\Delta\neq0$, computing (\ref{eq:O(t)O(t)}) requires
instead the contraction of a two dimensional network of width $\Delta+3$.
This is done by applying one
MPO at a time and truncating to the closest MPS with the given bond
$D$.  Since the network  is now finite
in both directions, this last phase of the contraction can be done
either in the spatial or in the time direction.


The success of the transverse approach will
depend on whether the transfer matrix of the evolved MPS 
has a non-degenerate dominant eigenvector 
which can be approximated by a MPS of reduced dimension.
Our implementation shows that the procedure achieves comparable 
results to the 
standard contraction~\cite{vidal07infinite} in a translationally invariant chain.
The transverse method offers the advantage of being 
applicable to dynamical situations in which 
translational symmetry is broken by a small number of sites, such as 
a chain with impurities, or a semi-infinite system, 
but it is also limited to short times.

However, there is a more efficient representation of the entanglement
in the transverse eigenvectors.
In the MPO representing the transfer matrix of the evolved MPS, tensors that
lie at the same distance from the center (occupied by the 
physical operator $O$ as in Fig.~\ref{fig:tn-b}) correspond to the 
same time step, coming from a certain term 
 and its adjoint in the Trotter decomposition.
We can group such pairs together in a new MPO by ``folding'' the original 
MPO (see Fig.~\ref{fig:tn-fold}).
The folding operation can be understood as performing 
the equivalent asymmetric contraction
$
\langle \Psi(t)|O|\Psi(t)\rangle=
\langle\Phi | \Bigl(O |\Psi(t)\rangle\otimes|\bar{\Psi}(t)\rangle \Bigr)
$
where $|\bar{\Psi}(t)\rangle$ is the complex conjugate of the evolved vector and
$|\Phi\rangle=\otimes_k\sum_{i_k=1}^d|i_k\bar{i}_k\rangle$ is the product of (unnormalized) maximally entangled pairs between
each site of the chain and its conjugate. 
In our scheme, the ket is now the tensor product of 
two tensor networks corresponding to $|\Psi(t)\rangle$ and its conjugate.
We may then group together each tensor in $|\Psi\rangle$ with the corresponding one
in $|\bar{\Psi}\rangle$, and define an effective tensor network of higher bond 
dimension  and physical dimension $d^2$, 
which can now be contracted using again the
transverse technique.

This folded transverse method allows us to explore the dynamics 
until much longer times than any other procedure. 
We may get some physical intuition for this improvement by looking at 
a single localized excitation that propagates freely with velocity $v$.
After time $t$, sites $x\pm vt$ in the evolved state become entangled. 
If we look instead at the transverse MPS obtained contracting the network from the right until $x+vt$, it is easy to see that all time sites are in a product, except for those corresponding to the instant $t$. 
These sites occupy symmetric positions around the center of the network,
so that folding groups them together in a single site which will be in a product state with all the rest. 

\begin{figure}[floatfix]
\begin{minipage}{\columnwidth}
\centering
\begin{psfrags}%
\psfragscanon%
\newcommand{\fntscl}{1.}
%
\psfrag{s01}[b][b][\fntscl]{\color[rgb]{0,0,0}\setlength{\tabcolsep}{0pt}\begin{tabular}{c}$\langle\sigma_{x}(t)\rangle$\end{tabular}}%
\psfrag{s05}[][][\fntscl]{\color[rgb]{0,0,0}\setlength{\tabcolsep}{0pt}\begin{tabular}{c} \end{tabular}}%
\psfrag{s06}[][][\fntscl]{\color[rgb]{0,0,0}\setlength{\tabcolsep}{0pt}\begin{tabular}{c} \end{tabular}}%
\psfrag{s07}[t][t][\fntscl]{\color[rgb]{0,0,0}\setlength{\tabcolsep}{0pt}\begin{tabular}{c}t\end{tabular}}%
\psfrag{s08}[b][b][\fntscl]{\color[rgb]{0,0,0}\setlength{\tabcolsep}{0pt}\begin{tabular}{c}$\epsilon_r$\end{tabular}}%
\psfrag{s16}[t][t][\fntscl]{\color[rgb]{0,0,0}\setlength{\tabcolsep}{0pt}\begin{tabular}{c}D=120\end{tabular}}%
\psfrag{s17}[b][b][\fntscl]{\color[rgb]{0,0,0}\setlength{\tabcolsep}{0pt}\begin{tabular}{c}D=60\end{tabular}}%
\psfrag{s18}[b][b][\fntscl]{\color[rgb]{0,0,0}\setlength{\tabcolsep}{0pt}\begin{tabular}{c}folded\\D=60\end{tabular}}%
\psfrag{s19}[b][b][\fntscl]{\color[rgb]{0,0,0}\setlength{\tabcolsep}{0pt}\begin{tabular}{c}folded\\D=120\end{tabular}}%
%
\psfrag{x01}[t][t][\fntscl]{0}%
\psfrag{x02}[t][t][\fntscl]{0.1}%
\psfrag{x03}[t][t][\fntscl]{0.2}%
\psfrag{x04}[t][t][\fntscl]{0.3}%
\psfrag{x05}[t][t][\fntscl]{0.4}%
\psfrag{x06}[t][t][\fntscl]{0.5}%
\psfrag{x07}[t][t][\fntscl]{0.6}%
\psfrag{x08}[t][t][\fntscl]{0.7}%
\psfrag{x09}[t][t][\fntscl]{0.8}%
\psfrag{x10}[t][t][\fntscl]{0.9}%
\psfrag{x11}[t][t][\fntscl]{1}%
\psfrag{x12}[t][t][\fntscl]{0}%
\psfrag{x13}[t][t][\fntscl]{5}%
\psfrag{x14}[t][t][\fntscl]{10}%
\psfrag{x15}[t][t][\fntscl]{0}%
\psfrag{x16}[t][t][\fntscl]{2}%
\psfrag{x17}[t][t][\fntscl]{4}%
\psfrag{x18}[t][t][\fntscl]{6}%
\psfrag{x19}[t][t][\fntscl]{8}%
\psfrag{x20}[t][t][\fntscl]{10}%
\psfrag{x21}[t][t][\fntscl]{12}%
%
\psfrag{v01}[r][r][\fntscl]{0}%
\psfrag{v02}[r][r][\fntscl]{0.1}%
\psfrag{v03}[r][r][\fntscl]{0.2}%
\psfrag{v04}[r][r][\fntscl]{0.3}%
\psfrag{v05}[r][r][\fntscl]{0.4}%
\psfrag{v06}[r][r][\fntscl]{0.5}%
\psfrag{v07}[r][r][\fntscl]{0.6}%
\psfrag{v08}[r][r][\fntscl]{0.7}%
\psfrag{v09}[r][r][\fntscl]{0.8}%
\psfrag{v10}[r][r][\fntscl]{0.9}%
\psfrag{v11}[r][r][\fntscl]{1}%
\psfrag{v12}[r][r][\fntscl]{$10^{-4}$}%
\psfrag{v13}[r][r][\fntscl]{$10^{-2}$}%
\psfrag{v14}[r][r][\fntscl]{$10^{0}$}%
\psfrag{v15}[r][r][\fntscl]{0.5}%
\psfrag{v16}[r][r][\fntscl]{1}%
\psfrag{v17}[r][r][\fntscl]{1.5}%
%
\includegraphics[width=.9\columnwidth,height=.5\columnwidth]{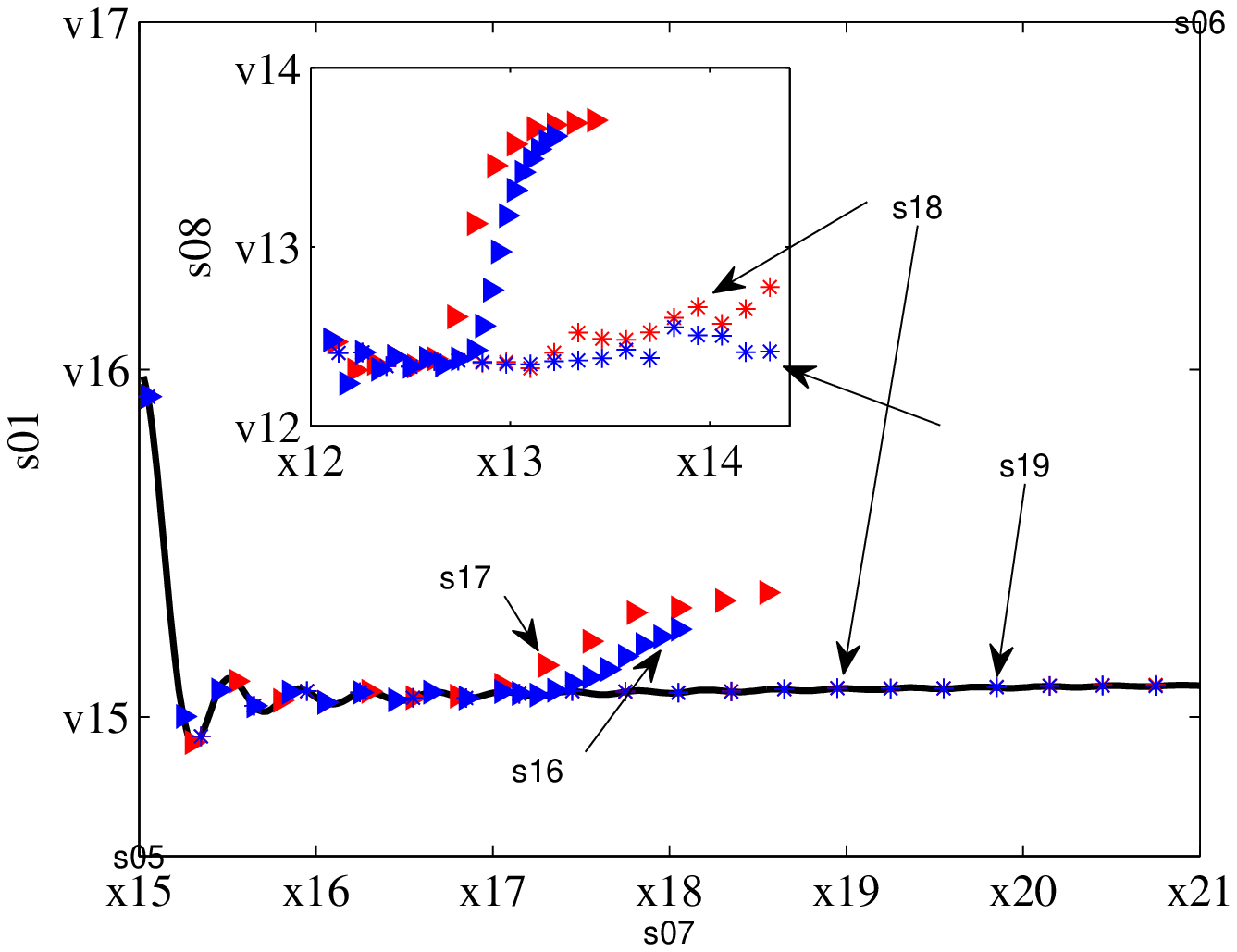}
\end{psfrags}
\end{minipage}
\\
\begin{minipage}{\columnwidth}
\centering
\begin{psfrags}%
\psfragscanon%
%
\psfrag{s02}[t][t]{\color[rgb]{0,0,0}\setlength{\tabcolsep}{0pt}\begin{tabular}{c}t\end{tabular}}%
\psfrag{s03}[b][b]{\color[rgb]{0,0,0}\setlength{\tabcolsep}{0pt}\begin{tabular}{c}$\langle\sigma_{x}(t)\rangle$\end{tabular}}%
\psfrag{s07}[][]{\color[rgb]{0,0,0}\setlength{\tabcolsep}{0pt}\begin{tabular}{c} \end{tabular}}%
\psfrag{s08}[][]{\color[rgb]{0,0,0}\setlength{\tabcolsep}{0pt}\begin{tabular}{c} \end{tabular}}%
\psfrag{s11}[][]{\color[rgb]{0,0,0}\setlength{\tabcolsep}{0pt}\begin{tabular}{c} \end{tabular}}%
\psfrag{s12}[][]{\color[rgb]{0,0,0}\setlength{\tabcolsep}{0pt}\begin{tabular}{c} \end{tabular}}%
\psfrag{s31}[t][t]{\color[rgb]{0,0,0}\setlength{\tabcolsep}{0pt}\begin{tabular}{c}folded\end{tabular}}%
\psfrag{s32}[t][t]{\color[rgb]{0,0,0}\setlength{\tabcolsep}{0pt}\begin{tabular}{c}D\end{tabular}}%
\psfrag{s37}[][]{\color[rgb]{0,0,0}\setlength{\tabcolsep}{0pt}\begin{tabular}{c}$\epsilon=10^{-6}$\end{tabular}}%
\psfrag{s38}[b][b]{\color[rgb]{0,0,0}\setlength{\tabcolsep}{0pt}\begin{tabular}{c}$\epsilon=10^{-8}$\end{tabular}}%
\psfrag{s39}[t][t]{\color[rgb]{0,0,0}\setlength{\tabcolsep}{0pt}\begin{tabular}{c}$\epsilon=10^{-2}$\end{tabular}}%
\psfrag{s40}[b][b]{\color[rgb]{0,0,0}\setlength{\tabcolsep}{0pt}\begin{tabular}{c}iTEBD\\D=64\end{tabular}}%
\psfrag{s41}[b][b]{\color[rgb]{0,0,0}\setlength{\tabcolsep}{0pt}\begin{tabular}{c}128\end{tabular}}%
\psfrag{s42}[b][b]{\color[rgb]{0,0,0}\setlength{\tabcolsep}{0pt}\begin{tabular}{c}256\end{tabular}}%
\psfrag{s43}[b][b]{\color[rgb]{0,0,0}\setlength{\tabcolsep}{0pt}\begin{tabular}{c}512\end{tabular}}%
\psfrag{s44}[b][b]{\color[rgb]{0,0,0}\setlength{\tabcolsep}{0pt}\begin{tabular}{c}1024\end{tabular}}%
\psfrag{s45}[b][b]{\color[rgb]{0,0,0}\setlength{\tabcolsep}{0pt}\begin{tabular}{c}$\epsilon=10^{-4}$\end{tabular}}%
%
\psfrag{x01}[t][t]{0}%
\psfrag{x02}[t][t]{0.1}%
\psfrag{x03}[t][t]{0.2}%
\psfrag{x04}[t][t]{0.3}%
\psfrag{x05}[t][t]{0.4}%
\psfrag{x06}[t][t]{0.5}%
\psfrag{x07}[t][t]{0.6}%
\psfrag{x08}[t][t]{0.7}%
\psfrag{x09}[t][t]{0.8}%
\psfrag{x10}[t][t]{0.9}%
\psfrag{x11}[t][t]{1}%
\psfrag{x12}[t][t]{0}%
\psfrag{x13}[t][t]{5}%
\psfrag{x14}[t][t]{10}%
\psfrag{x15}[t][t]{0}%
\psfrag{x16}[t][t]{2}%
\psfrag{x17}[t][t]{4}%
\psfrag{x18}[t][t]{6}%
\psfrag{x19}[t][t]{8}%
\psfrag{x20}[t][t]{10}%
%
\psfrag{v01}[r][r]{0}%
\psfrag{v02}[r][r]{0.2}%
\psfrag{v03}[r][r]{0.4}%
\psfrag{v04}[r][r]{0.6}%
\psfrag{v05}[r][r]{0.8}%
\psfrag{v06}[r][r]{1}%
\psfrag{v07}[r][r]{$10^{1}$}%
\psfrag{v08}[r][r]{$10^{2}$}%
\psfrag{v09}[r][r]{0.4}%
\psfrag{v10}[r][r]{0.6}%
\psfrag{v11}[r][r]{0.8}%
\psfrag{v12}[r][r]{1}%
%
\includegraphics[width=.9\columnwidth,height=.5\columnwidth]{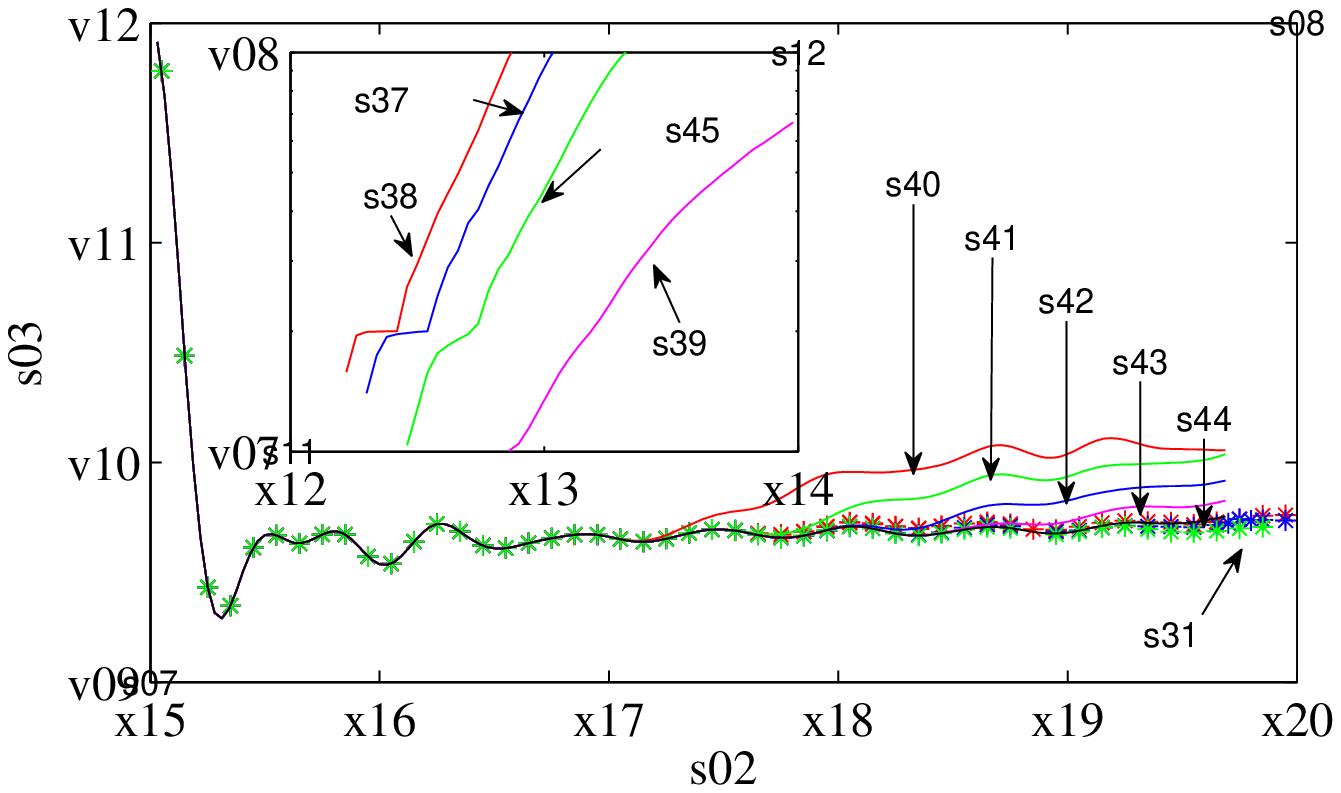}
\end{psfrags}
\end{minipage}
\caption{Magnetization as a function of time.
For the Ising model (up) results for the transverse method (triangles) are compared
to the folded version (stars) for $D=60,$ $120$. The relative error
with respect to the exact result (solid line) is shown in the inset. 
For the non-integrable model (down), results with the folded approach for D=60 (red), 120 (blue), 240 (green) are compared to
those of iTEBD (solid lines) for increasing values of $D$. In the inset, the
required value of $D$ as a function of time, for different
levels of accuracy.
}
\label{fig:Ising}
\label{fig:IsingP}
\end{figure}

As a first benchmark for the new method,
we simulate the dynamics of states far from equilibrium
under the Ising Hamiltonian (\ref{eq:Ising}) with uniform magnetic field $g$.
The initial state
$|\Psi_0\rangle=\otimes_i\frac{1}{\sqrt{2}}(|0\rangle_i+|1\rangle_i)$
is evolved with a constant Hamiltonian and the results of the
transverse method with and without folding are compared to the exact results
(Fig.~\ref{fig:Ising}).  
For very short times the Trotter error dominates in both methods. 
However, while for the
transverse procedure (as for iTEBD) truncation error becomes
soon dominant, and the results deviate abruptly from the exact
solution, the accuracy of the folded version is
maintained for much longer times.

To test the method on a more general problem, we repeat the test for a non-integrable Hamiltonian,
$
H=-\left(\sum_i\sigma_z^i \sigma_z^{i+1}+g \sigma_x^i+h \sigma_z^i  \right ).
$
For this case there are no exact results, but we may compare the 
folded computation to the iTEBD simulations with a similar Trotter 
error (Fig.~\ref{fig:IsingP}).
Again we check that the accuracy of the folded procedure for comparable bond dimension
reaches much longer times. Moreover, remarkably enough,
even when the results from the folded method start deviating 
(from those to which iTEBD converges for large $D$),
they do so in a smooth way, so that, in contrast to other procedures, they 
continue to qualitatively describe the evolution for long times.

This can be seen in a more precise way by looking at the truncation error. 
At a certain time, this can be estimated by 
looking at the error in the right eigenvector for a given bond dimension
with respect to the best eigenvector obtained, i.e. that for the highest $D$.
If we plot (Fig.~\ref{fig:IsingP}) the bond dimension
required to achieve a fixed truncation error, we observe that, although 
the $D$ required for a high precision grows exponentially with time, 
with a relatively low bond $D<100$ a qualitative description of the dynamics is
reproduced, that lies within $1\%$ of the exact solution well beyond times $t>10$.



From the discussion above the transverse method, combined with the
folding technique, represents a very promising tool for the dynamical studies
of one dimensional systems.  The first results show the
applicability of the method even to non-integrable systems, allowing
the simulation of longer evolution times than any other technique, and
a qualitative description of the dynamics until even later.  This
opens the door for the study of physical problems not accessible until
now for numerical methods, including the dynamics of phase
transitions, out-of-equilibrium states and thermalization problems.
The present formalism might also prove very valuable in the context of
extracting spectral information for quantum impurity problems, the 
central problem in dynamical mean field theory. The big
advantage of our method is that we can deal with real frequencies, and
no analytic continuation from imaginary frequencies is needed as in
the case of Monte Carlo simulations. In contrast, the main limitation
would be its exclusive applicability to one dimensional systems.
Finally, although the method has been
described for infinite chains, it is easy to adapt the technique
for the dynamical study of finite systems.

\paragraph*{Note: } 
An independent derivation in~\cite{huebener09ctn}, in the context of
concatenated tensor network states, lead to a similar network to describe 
time-evolved states.

\acknowledgments
We acknowledge D. P\'erez-Garc\'{\i}a and J. J. Garc\'{\i}a-Ripoll 
for discussions
and T. Nishino for pointing out Ref.~\cite{rommer99imp}.
This work was supported 
by DFG through Excellence Cluster MAP 
and FOR 635, and by EU project SCALA.


\end{document}